\documentstyle[12pt,epsfig]{article}

\def\tou#1{{\lower1.2ex\hbox{$\longrightarrow$}\atop
        {\lower-.7ex\hbox{$\scriptscriptstyle #1 $}}}}
\def\lsim{{\lower1.2ex\hbox{$<$}\atop
        {\lower-.7ex\hbox{$\sim$}}}}
\def\gsim{{\lower1.2ex\hbox{$>$}\atop
        {\lower-.7ex\hbox{$\sim$}}}}

\begin{document}

\begin{titlepage}
\rightline {hep-th/9311175 \  \  \  \   }

\rightline {Si-93-06 \  \  \  \   }

\vskip 3truecm
\centerline {\Large\bf Duality of a
 Generalized Gauge Invariant Ising Model}
\vskip .4cm
\centerline {\Large\bf on Random Surfaces\footnote{
  Work supported in part by the Deutsche Forschungsgemeinschaft,
  Az. Schu95/7-2}}

\vskip 2.0truecm

\centerline{\bf Z.B. Li\footnote{Alexander-von-Humboldt-fellow.
On leave of absence from the Zhongshan University, 510275 Guangzhou,
P.R. China},
B. Zheng and L. Sch\"ulke}

\vskip 0.7truecm

\centerline{Universit\"at-GH Siegen, D-57068 Siegen, Germany}

\vskip 3truecm

\abstract
A generalized gauge invariant Ising model on random surfaces
with non-trivial topology
is proposed and investigated with the dual transformation.
It is proved that the model is self-dual in case of a
self-dual lattice. In special cases the model reduces to the
known solvable Ising-type models.

\end{titlepage}

\section {Introduction}

In recent years great interest has turned to the non-critical
string theory with central charge $C\leq1$, which was claimed
to be non-perturbatively solvable by relating
its dual model to certain matrix representations.

The string theory with $C=1/2$ can be viewed as the Ising model
on random surfaces.
In the non-perturbative
calculation, the continuum surface is tessellated by polygons
with $p$ edges. Then
through the dual transformation, the Ising model is related to its dual
(Ising) model on the $\phi^p$ lattice.
The latter has a two-matrix representation which is
non-perturbatively  solvable around the double scaling point
\cite{GRO90,BRE90,CRN90}.

The dual tranformation, however, is not straightforward for non-planar
lattices because the high-temperature expansion contains
non-boundary one-cycles. As early as 1971, Wegner
\cite{WEG71} pointed out that the
traditional duality relation is valid only if the completeness
condition is fulfilled. For two dimensional lattices
Wegner's completeness condition means nothing but the
planar lattice.

Recently a generalized duality relation valid for both planar and
non-planar
lattices has been obtained \cite{LI92,LI92a}. The remarkable
fact is that the dual model of an Ising model
is not simply an Ising model, as one
implicitly supposed in the two-matrix representation of the string
theory with $C=1/2$.
It has also been shown that the model, whose dual model is an Ising
model coupled to an external field, turns out to be
a gauge invariant Ising model where non-boundary one-cycles do not
appear due to the gauge invariance.
This was first suggested by Wegner \cite{WEG71}
and Balian et al. \cite{BAL75} for a regular
planar lattice.
For what has recently been done,
the exact statement should, therefore, be
that it is the gauge invariant Ising model on random surfaces that
has the solvable two-matrix representation.
Whether the gauge invariant Ising model and the Ising model
coupled to an external field belong to the same
universality class remains open.\footnote{
One may directly discretize the continuum
theory by putting the spins on the dual lattice, but in this case
one also has to know whether this discretized model is
in the same universality class of the original one or not.}

All these indicate that {\em the gauge field in two dimensions is
far from trivial, especially in the case of non-planar lattices}.
It is therefore interesting to investigate Ising-type models
on random surfaces with gauge interaction and to study
the topological effects of the gauge field.

On the other hand, as is well known, the self-duality is one of the
most attractive points of the Ising model, which help us to understand
the phase structure. The self-duality is in general valid only
when the lattice is self-dual. For the planar random surfaces
one may get the self-duality by summing up all the lattices.
This has recently been discussed \cite{ALV90}.
On non-planar lattices the situation is more complicated.
The self-duality is spoiled by the non-trivial
topology of the lattice even if the lattice is self-dual.
{\em What is then a self-dual
model on non-planar random surfaces? }

Furthermore the dual transformation itself is also useful in studying
the relation between two Ising-type models \cite{SAV80}.
In two dimensions the topological properties of
the surfaces are closely related to
boundary conditions. Therefore it is important to know how
the dual transformation goes for topologically different lattices.

In order to tackle these problems, in this paper
the simple gauge invariant Ising model \cite{WEG71,BAL75}
is generalized such that two Ising-spins interact with a gauge field.
It will be seen that this model is
self-dual even for a non-planar lattice if the lattice is self-dual,
because the gauge field kills the non-boundary one-cycles in
the high-temperature representation.
This is the simplest self-dual Ising-type model on the lattice with
non-trivial topology.
Under certain limits the model reduces to relatively
simpler ones, which are already known to be physically
interesting and important.
We can demonstrate the calculation of
Wilson loops and show how the topology
plays its role.

In the next section the dual transformation will be carried out.
Several limiting cases
are  considered in section~3.
Finally some discussions follow.

\section {A self-dual model}

Let us consider a system on a random lattice $L$ with two
sets of Ising spins
and a $Z_2$ gauge field, whose partition function is
\begin{equation}\begin{array}{c}{\displaystyle
Z=\sum_{ \{S_i,S^{'}_i,U_{ij}\}   }
     \exp \left\{\sum_{<ij>\in L^1 } \left(JU_{ij} S_i S_j
    +J^{'} U_{ij} S^{'}_i S^{'}_j\right) \right. \hspace*{2cm}
}\\{\displaystyle
  \hspace*{4cm} \left. +\sum_{m\in L^2 } \beta U_m
    +\sum_{i\in L^0 } \lambda  S_i S^{'}_i \right\}
\label{parz}
}\end{array}\end{equation}
where $L^0$,$L^1$ and $L^2$ are respectively the set of
sites, bonds and plaquettes; $S_i=\pm 1$
and $S^{'}_i=\pm 1$ are the spin variables on a
site, $U_{ij}=\pm 1$ is the gauge field on a bond and $U_m$ is
the plaquette variable defined as the product of the gauge fields
along the boundary of the plaquette.
This model looks like the {\em `spin-Schwinger model'}.
$\sum_{i\in L^0 } \lambda  S_i S^{'}_i$
corresponds to the `mass' term.

Here we should  mention, as in the case of the simple gauge invariant
Ising model \cite{BAL75},  in the `vacuum sector' the
spins $\{S^{'}_i\}$ can be
gauged away by the following tranformation
\begin{equation}
{\tilde U}_{ij}= U_{ij} S^{'}_i S^{'}_j, \quad
{\tilde S}_i= S_i S^{'}_i
\end{equation}. However, for the convenience of discussions
we will keep it because it preserves explicitly the gauge
invariance of the system.

For a given random lattice $L$, one can construct a dual lattice
$L^D$, and a  model on it with the partition function
\begin{equation}\begin{array}{c}{\displaystyle
Z_D=\sum_{\{\tau_m,\tau^{'}_m,V_{mn}\}}
     \exp \left\{\sum_{<mn>\in L_D^1 } \left(J_D V_{mn}
                \tau_m \tau_n
    +J_D^{'} V_{mn} \tau^{'}_m \tau^{'}_n\right)\right.  \hspace*{1cm}

}\\{\displaystyle
  \hspace*{3cm}   \left. +\sum_{i\in L_D^2 } \beta_D V_i
    +\sum_{m\in L_D^0 } \lambda_D  \tau_m \tau^{'}_m\right\}
\label{parzd}
}\end{array}\end{equation}
where $L_D^0$,$L_D^1$ and $L_D^2$ are respectively the set of
dual sites, dual bonds and dual plaquettes;
$\tau_m, \tau^{'}_m=\pm1$ are dual spin variables on
dual sites, $V_{mn}=\pm 1$ is the dual gauge field
on a dual bond and $V_i$ is
the dual plaquette variable defined as
the product of the dual gauge fields
along the boundary of the dual plaquette.

Following the standard procedure for the dual transformation,
we will prove that the model with $Z$ is dual to the one with $Z_D$.
Rigorously speaking,
\begin{equation}
Z=c Z_D,\quad c=2^{3(N-P)/2} (\sinh   2J  \sinh   \,2J^{'})^{B/2}
      (\sinh   2 \beta)^{P/2} (\sinh   2 \lambda )^ {N/2}
\label{dualz}
\end{equation}
if
\begin{equation}
\beta_D=\frac{1}{2}\,\ln\,(\coth   \lambda), \quad
\lambda_D=\frac{1}{2}\,\ln\,(\coth   \beta)
\label{dualbl}
\end{equation}
and either
\begin{equation}
J_D=\frac{1}{2}\,\ln\,(\coth   J),\quad J^{'}_D=\frac{1}{2}\,\ln\,(\coth
J^{'})
\label{dualjs}
\end{equation}
or
\begin{equation}
J_D=\frac{1}{2}\,\ln\,(\coth   J^{'}), \quad J^{'}_D=\frac{1}{2}\,\ln\,(\coth
J) ,
\label{dualja}
\end{equation}
where $B$, $N$ and $P$ are respectively the numbers of the bonds,
sites and plaqettes on the original lattice.

Note that if $L\neq L_D$, $Z$ and $Z_D$ are respectively partition functions
for two different models,
which are dual each other.
But in case of $L= L_D$, these two model become the same, i.e.
we have a self-dual model.
Then the dual relations in (\ref{dualbl}-\ref{dualja})
may give some insight in
the  phase structure.
{}From the dual relation (\ref{dualbl}) we know that
the critical points appear always by pair in the
$(\beta,\lambda)$ plane. It is, however,
not possible to locate the critical point even when it is
assumed to be unique, as in the case of the simple
Ising model. From the dual relations (\ref{dualjs}-\ref{dualja})
one can see that the critical points are grouped in four
in the $(J,J^{'})$ plane.
If there is only a unique critical point in the $(J,J^{'})$ plane,
it is $(J^{*},J^{*})$ with $J^{*}=\frac{1}{2}\,\ln\,(\coth   J^{*})$.
To make full use of the dual relations further investigations are needed.

In order to get the duality relations (\ref{dualz}-\ref{dualja}),
we simply write down the high-temperature representation for $Z$
and the low-temperature representation for $Z_D$ and  compare the two
expressions.

\subsection{The high-temperature representation of $Z$}

Noting that $S_i$, $S^{'}_i$ and $U_{ij}$ all take values $\pm 1$,
the partition function $Z$ in Eq.(\ref{parz}), can be written as

\begin{equation}\begin{array}{c}{\displaystyle
Z=\sum_{ \{S_i,S^{'}_i,U_{ij}\}   }
      \prod_{<ij>\in L^1 } \left(\cosh  J+U_{ij} S_i S_j\sinh  J\right)
     \prod_{<ij>\in L^1 } \left(\cosh J^{'}+U_{ij} S^{'}_i S^{'}_j\sinh
J^{'}\right)
}\\{\displaystyle \hspace*{2cm}
      \prod_{i\in L^0 } \left(\cosh \lambda+S_i S^{'}_i\sinh  \lambda\right)
     \exp\left\{\sum_{m\in L^2 } \beta U_m \right\}
}\end{array}\end{equation}
The product $\prod_{<ij>\in L^1 } (\cosh J+U_{ij} S_i S_j\sinh  J)$
can be expanded into $2^B$ terms. For a given term, a bond
$<ij>$ contributes a factor of either $\cosh J$ or $U_{ij} S_i S_j\sinh  J$.
Let us mark the bond in case of $U_{ij} S_i S_j\sinh  J$.
All the marked bonds for this given term assemble
as a one-chain $\gamma$
on the lattice. It is easy to see that this is a one-to-one
correspondence between all the terms in the expansion
and all the one-chains on the lattice.
Similarly $\prod_{<ij>\in L^1 }(\cosh J^{'}+U_{ij} S^{'}_i S^{'}_j\sinh
J^{'})$
can be expanded according to another set of all one-chains
$\gamma^{'}$. For $\prod_{i\in L^0 } (\cosh \lambda+S_i S^{'}_i\sinh  \lambda)$
 we mark
the site $i$
if a term in the expansion picks up $S_i S^{'}_i\sinh  \lambda$.
All the marked sites in the term form
a zero-chain $\theta$. This is also a one-to-one correspondence.
After performing the summation over $\{S_i\}$ and $\{S^{'}_i\}$,
we get

\begin{equation}\begin{array}{c}{\displaystyle
Z=2^{2N}\sum_{ \{U_{ij}\}   } \sum_{\gamma,\gamma^{'}\in C^1}
       \sum_{\theta\in C^0} \;\;
      \prod_{<ij>\in \gamma } \sinh  J
      \prod_{<ij> \notin\gamma} \cosh J
      \prod_{<ij>\in \gamma^{'} }  \sinh  J^{'}
}\\{\displaystyle
      \prod_{<ij> \notin\gamma^{'}} \cosh J^{'}
     \prod_{i\in \theta } \sinh  \lambda
     \prod_{i \notin\theta } \cosh \lambda
     \prod_{<ij>\in \gamma_0 } U_{ij}\;
     \exp\left\{\sum_{m\in L^2 } \beta U_m \right\}
}\end{array}\end{equation}
where $C^0$ and $C^1$ are the sets of zero-chains and one-chains
respectively, $\gamma_0=\gamma+\gamma^{'}$,
and
\begin{equation}
\theta+\partial\gamma=0, \quad
\theta+\partial\gamma^{'}=0
\label{constraint}
\end{equation}
Here $\partial\gamma$
and $\partial\gamma^{'}$ denote respectively the boundaries of
$\gamma$ and$\gamma^{'}$.
The equation (\ref{constraint} ) simply means that $\gamma$ and
$\gamma^{'}$ are connected by $\theta$ resulting
that $\gamma_0$ must be a one-cycle.

Here we should stress that up to now the gauge invariance has
not been considered and $\gamma_0$ can be
either a boundary one-cycle, which circles a certain area of
the surface, or a non-boundary one-cycle, which
bounds no area.
This is shown in Fig~\ref{fig1}.

     \begin{figure}[t]\centering


\vspace{-10.cm}

\setlength{\unitlength}{1cm}
\begin{picture}(10,10)
  \put (-1,-0.5){ $\gamma+\gamma'=\gamma_0$ }
  \put (7.7,0.65){ $\gamma+\gamma'=\gamma_0$ }
  \put (2.2,2.45){$\gamma$}
  \put (6.8,5.2){$\gamma$}
  \put (8.2,3.23){$\gamma'$}
  \put (2.9,2){$\gamma'$}
  \put (4.3, 5.35){$\theta$}
  \put (4.65,1.87){$\theta$}
\end{picture}
\vspace{1cm}
\caption{
This is a handle of the random surface.
The $\gamma_0$ on the right side is a boundary one-cycle.
The $\gamma_0$ on the left is a non-boundary one-cycle
}
\label{fig1}
     \end{figure}

Similarly one can carry out the summation
over gauge fields. If $\gamma_0$ is a boundary one-cycle,

\begin{equation}\begin{array}{c}{\displaystyle
  \sum_{ \{U_{ij}\}   }
     \prod_{<ij>\in \gamma_0 } U_{ij}
     \exp\left\{\sum_{m\in L^2 } \beta U_m \right\} =  \hspace*{5cm}
}\\ {\displaystyle \hspace{1cm}
     2^B\left\{ \prod_{m\in \delta\gamma_0 } \sinh  \beta
     \prod_{m \notin\delta\gamma_0 } \cosh \beta
     +\prod_{m\in \delta\gamma_0 } \cosh \beta
       \prod_{m \notin\delta\gamma_0 } \sinh  \beta \right\}
}\end{array}\end{equation}
where $\delta\gamma_0$ is the set of all plaquettes
with boundary $\gamma_0$. The two terms on the
right-hand side represent the two ways to fill up $\gamma_0$,
`inside' and `outside'.
If $\gamma_0$ is a non-boundary one-cycle,
the summation over gauge fields gives zero.
Therefore

\begin{equation}\begin{array}{c}{\displaystyle
Z=2^{B+2N} \sum_{\gamma\in C^1}
       \sum_{\gamma_0\in \Gamma_0}
      \prod_{<ij>\in \gamma } \sinh  J
      \prod_{<ij>\notin \gamma} \cosh J\,  \hspace*{2cm}
}\\{\displaystyle \hspace*{2cm}
       \prod_{<ij>\in \gamma+\gamma_0 } \sinh  J^{'}
      \prod_{<ij> \notin\gamma+\gamma_0} \cosh J^{'}
     \prod_{i\in \partial\gamma } \sinh  \lambda
     \prod_{i \notin\partial\gamma } \cosh \lambda
}\\{\displaystyle  \hspace*{2cm}
     \left(\prod_{m\in \delta\gamma_0 } \sinh  \beta
     \prod_{m \notin\delta\gamma_0 } \cosh \beta
     +\prod_{m\in \delta\gamma_0 } \cosh \beta
       \prod_{m \notin\delta\gamma_0 } \sinh  \beta\right)
\label{high}
}\end{array}\end{equation}
where $\Gamma_0$ is the set of all boundary one-cycles.
This is the so-called high-temperature representation of $Z$.
{\em The gauge field damps the non-boundary
one-cycles induced by the non-trivial topology of the lattice.}

\subsection{The low-temperature representation}

The configurations of the dual gauge fields can graphically be
represented by all the one-chains $C^1$ on the original lattice.
Let us denote the dual bond of $<mn>$ by $<mn>^*$.
For a given configuration, the corresponding one-chain $\gamma$
is obtained as follows: if $V_{mn}=-1$, draw a line on the
bond $<mn>^{*}$ on the original lattice;
if $V_{mn}=1$, do nothing. It is not difficult to see that on the
boundary of $\gamma$ the dual plaquette variables are
$-1$ and otherwise $+1$. This is shown in Fig~\ref{fig2}.

     \begin{figure}[t]\centering


\vspace{-5cm}

\setlength{\unitlength}{1cm}
\begin{picture}(10,5)
\put(5.6,  4.7){$V_{m\,n}$}
\put(2.0,  1.0){$\gamma$}
\end{picture}

\caption{
This is a part of the original lattice and the dual lattice.
Here the circles and the dotted lines represent the dual sites and
dual bonds.
The real dots and lines are the sites and bonds on the original lattice.
$V_{mn}$ is the dual gauge field.
}
\label{fig2}
\end{figure}

Therefore $Z_D$ can be written as
\begin{equation}
\begin{array}{c}{\displaystyle
Z_D=\sum_{ \{\tau_m,\tau^{'}_m\}   } \sum_{ \gamma\in C^1   }
      \prod_{<mn>^{*}\in \gamma }
     \exp\left\{-J_D \tau_m\tau_n-J^{'}_D \tau^{'}_m\tau^{'}_n\right\}
\hspace{3cm} }
\\  \hspace{2cm}
{\displaystyle
 \prod_{<mn>^{*} \notin\gamma }
     \exp\left\{J_D \tau_m\tau_n+J_D \tau^{'}_m\tau^{'}_n\right\} }\\
{\displaystyle  \hspace{2cm}
       \prod_{i\in \partial\gamma } e^{-\beta_D}
        \prod_{i \notin\partial\gamma } e^{\beta_D}
     \exp\left\{\sum_{m\in L^0_D } \lambda_D \tau_m\tau^{'}_m \right\}
}\end{array}
\end{equation}

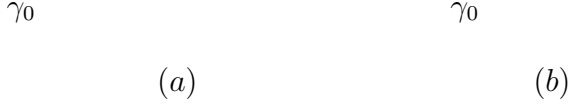
\begin{figure}[t]\centering

  \begin{tabular}{cc}
  \end{tabular}
\vspace{-5cm}

\setlength{\unitlength}{1cm}
\begin{picture}(11,5)
\put(5,4.5){$\tau_m$}
\put(10.9,4.5){$\tau_m$}
\put(0.5,0){$\gamma_0$}
\put(6.4,0){$\gamma_0$}
\put(2.5,-1.0){$(a)$}
\put(7.5,-1.0){$(b)$}
\end{picture}

\vspace{1.4truecm}
\caption{
(a) and (b) are the two configurations of dual spins $\{\tau_m\}$
corresponding to the boundary one-cycle $\gamma_0$.}
\label{fig3}
     \end{figure}

For  a given dual spin configuration $\{\tau_m\}$, as shown in
Fig.~\ref{fig3}
we can draw a boundary
one-cycle $\gamma_0$ on the original lattice,
such that all the dual spins $\tau_m$ inside $\gamma_0$
have the same sign. This is a two-to-one correspondence.
In other words, for a given boundary one-cycle, one can assign
either positive spins or negative spins inside it.
For $\{\tau^{'}_m\}$ we get similarly a correspondence to
$\gamma^{'}_0$.
Then we have

\begin{equation}\begin{array}{c}{\displaystyle
Z_D= 2 \sum_{\gamma\in C^1}
       \sum_{\gamma_0,\gamma^{'}_0\in \Gamma_0}
      \prod_{<mn>^{*}\in \gamma+\gamma_0 } e^{-J_D}
      \prod_{<mn>^* \notin\gamma+\gamma_0} e^{J_D}  \hspace*{2cm}
}\\{\displaystyle
       \prod_{<mn>^*\in \gamma+\gamma^{'}_0 } e^{-J^{'}_D}
      \prod_{<mn>^* \notin\gamma+\gamma^{'}_0} e^{J^{'}_D}
     \prod_{i\in \partial\gamma} e^{-\beta_D}
     \prod_{i \notin\partial\gamma } e^{\beta_D}
}\\{\displaystyle  \hspace*{2cm}
   \left(\prod_{m\in \delta(\gamma_0+\gamma^{'}_0)  } e^{-\lambda_D}
   \prod_{m \notin\delta(\gamma_0+\gamma^{'}_0)  } e^{\lambda_D}
   +\prod_{m\in \delta(\gamma_0+\gamma^{'}_0)  } e^{\lambda_D}
    \prod_{m \notin\delta(\gamma_0+\gamma^{'}_0)  } e^{-\lambda_D}\right)
 \hspace*{2cm}
}\end{array}\end{equation}
After renaming $\gamma+\gamma_0$ as $\gamma$ and
$\gamma_0+\gamma^{'}_0$  as $\gamma_0$,

\begin{equation}\begin{array}{c}{\displaystyle
Z_D= 2^P \sum_{\gamma\in C^1}
       \sum_{\gamma_0\in \Gamma_0}
      \prod_{<mn>^{*}\in \gamma } e^{-J_D}
      \prod_{<mn>^* \notin\gamma} e^{J_D}  \hspace*{2cm}
}\\{\displaystyle \hspace*{2cm}
       \prod_{<mn>^*\in \gamma+\gamma_0 } e^{-J^{'}_D}
      \prod_{<mn>^* \notin\gamma+\gamma_0} e^{J^{'}_D}
       \prod_{i\in \partial\gamma } e^{-\beta_D}
     \prod_{i \notin\partial\gamma } e^{\beta_D}
}\\{\displaystyle  \hspace*{2cm}
      \left(\prod_{m\in \delta\gamma_0 } e^{-\lambda_D}
     \prod_{m \notin\delta\gamma_0 } e^{\lambda_D}
     +\prod_{m\in \delta\gamma_0 } e^{\lambda_D}
       \prod_{m \notin\delta\gamma_0 } e^{-\lambda_D}\right)
\label{low}
}\end{array}\end{equation}
This is the low-temperature representation of $Z_D$.
Comparing (\ref{high}) and (\ref{low}), we can easily arrive at the
dual relations (\ref{dualz}-\ref{dualja}).

\section {Some limiting cases}

For some special limiting cases the model reduces to
known ones which are solvable.
Then the dual relations in (\ref{dualbl}-\ref{dualja})
may be regarded as the dual relations between two different models.
Sometimes the reduction of the model is non-trivial and interesting.

\subsection{The limit $\lambda\rightarrow 0$}

{}From eq.(\ref{dualbl}) we have ${\beta_D\rightarrow \infty}$.
All the dual plaquettes
will be frozen to $V_i=1$. This is a constraint for the dual
gauge fields $\{V_{mn}\}$.
In the language of cohomololgy theory, $\{V_i\}$
are two-forms, gauge fields $\{V_{mn}\}$ are one-forms. Due to
the constraint, $\{V_{mn}\}$ reduce to one-cocycles.

The {\em exact} one-cocycles can be written as
\begin{equation}
V^{(exact)}_{mn}=\eta_m\eta_n
\end{equation}
where $\eta_m,\eta^{'}_m=\pm$ are also
spin variables. That is, $\{V^{(exact)}_{mn}\}$
are pure gauges. It is known that all one-cocycles
can be classified into $\alpha=2^{2g}$ cohomology classes
by the equivalent
relation: $\{V_{mn}\}$ and $\{V^{'}_{mn}\}$ are cohomologically
equivalent
if $\{V_{mn}V^{'}_{mn}\}$ is an {\em exact} one-cocycle;
otherwise, they are not cohomologically equivalent.

In our case, the cohomological equivalence
is just the same as the local gauge equivalence.
Cohomology classes are
trajectories of  gauge fields under the local gauge
transformation. If one can find an arbitrarily specified one-cocycle
for each cohomology class, say $\{{\tilde V}^k_{mn}\}$,
then the general solution of the constraint
is
\begin{equation}
V_{mn}= {\tilde V}^k_{mn}\eta_m\eta_n, \quad k=0,1,...,\alpha-1
\end{equation}
The specified one-cocycle of the $k$-$th$ cohomology class can be
constructed from a one-cycle in the $k$-$th$ homology class.
Let $\gamma_k$ be an arbitrarily specified one-cycle of the $k$-$th$ homology
class on the original lattice, one can define
\begin{equation}
{\tilde V}^k_{mn}= {\tilde V}_{mn} (\gamma_k) \equiv
\cases{
-1   &  $<mn>^{*}\in\gamma_k$\cr
\ 1  &  $<mn>^{*}\notin\gamma_k$\cr}
\end{equation}
Therefore the dual partition function in the limit
$\lambda\rightarrow 0$ is

\begin{equation}\begin{array}{c}{\displaystyle
Z_D=\frac{1}{2}\sum^{\alpha-1}_{k=0}
      \sum_{\{\tau_m,\tau^{'}_m,\eta_{m}\}}
     \exp \left\{\sum_{<mn>\in L_D^1 } {\tilde V}_{mn} (\gamma_k) \eta_m\eta_n
            (J_D  \tau_m \tau_n
    +J_D^{'} \tau^{'}_m \tau^{'}_n) \right.  \hspace*{.5cm}
}\\{\displaystyle
 \hspace*{8cm}
      \left. +\sum_{m\in L_D^0 } \lambda_D  \tau_m \tau^{'}_m\right\}
}\end{array}\end{equation}
Here the factor $1/2$ arises due to the fact that $\{\eta_m\}$
and $\{-\eta_m\}$ correspond to the same gauge field ${V}_{mn}$.
One can redefine the spin variables
$\tau_m \eta_m \rightarrow \tau_m$ and
$\tau^{'}_m \eta_m \rightarrow \tau^{'}_m$,
then the dual relation $Z=c\;Z_D$, (\ref{dualz}), reduces to

\begin{equation}\begin{array}{c}{\displaystyle
  \sum_{ \{S_i,S^{'}_i,U_{ij}\}   }
     \exp \left\{\sum_{<ij>\in L^1 } \left(JU_{ij} S_i S_j
    +J^{'} U_{ij} S^{'}_i S^{'}_j\right)
    +\sum_{m\in L^2 } \beta U_m\right\} =  \hspace*{2cm}
}\\{\displaystyle
 \hspace*{2cm}
 c\sum^{\alpha-1}_{k=0}
      \sum_{\{\tau_m,\tau^{'}_m\}}
     \exp \left\{\sum_{<mn>\in L_D^1 } {\tilde V}_{mn} (\gamma_k)
            \left(J_D  \tau_m \tau_n
    +J_D^{'} \tau^{'}_m \tau^{'}_n\right) \right.
}\\{\displaystyle \hspace*{4cm}
    \left. +\sum_{m\in L_D^0 } \lambda_D  \tau_m \tau^{'}_m\right\}
}\end{array}\end{equation}
with $c= 2^{2N-1-P/2}(\sinh   2J  \sinh   \,2J^{'})^{B/2}
      (\sinh   2 \beta)^{P/2}$ .The left-hand side is
a `massless' two-spin system
with gauge interaction.
The right-hand side is its dual model,
whose partition function contains several terms
due to the non-trivial topology of the lattice,
 of which each describes a `massive' two-spin
system with an anti-ferromagnetic chain.
Both models are very interesting.
The techniques of matrix representation may also be applied
for solving them.

For the planar lattice we have $\alpha=1$ $(g=0)$. When
$\lambda_D \rightarrow \infty$, the dual model reduces to
a simple Ising model and the phase transition
appears at $J_D+J^{'}_D=J^{*}$. This is a critical line.
If one comes back the original model, the critical points
lie on the curve $1/2\,\ln\,\coth    J +1/2\,\ln\,\coth    J^{'}=J^{*}$
when $\beta$ also goes to zero.

\subsection{The limit $\lambda\rightarrow 0,\; \beta \rightarrow \infty$}

Let us further reduce the model in last subsection
by taking the limit $\beta \rightarrow \infty$, which implies
$\lambda=0$, see eq.(\ref{dualbl}).
Now the model becomes self-dual and the self-dual relation is

\begin{equation}\begin{array}{c}{\displaystyle
   \sum^{\alpha-1}_{k=0}
  \sum_{ \{S_i,S^{'}_i\}   }
     \exp \left\{\sum_{<ij>\in L^1 }
     {\tilde U}_{ij} \left(\gamma^{D}_k)(J S_i S_j
    +J^{'} S^{'}_i S^{'}_j\right)\right\} =   \hspace*{2cm}
}\\{\displaystyle
 \hspace*{2cm}
    c\sum^{\alpha-1}_{k=0}
      \sum_{\{\tau_m,\tau^{'}_m\}}
     \exp \left\{\sum_{<mn>\in L_D^1 } {\tilde V}_{mn} (\gamma_k)
            \left(J_D  \tau_m \tau_n
    +J_D^{'} \tau^{'}_m \tau^{'}_n\right)\right\}
}\end{array}\end{equation}
where $c= 2^{N-P}(\sinh   2J  \sinh   \,2J^{'})^{B/2}$,
\begin{equation}
{\tilde U}_{ij} (\gamma^{D}_k) \equiv
\cases{
-1 &  $<ij>^{*}\in\gamma^{D}_k$\cr
\ 1  &  $<ij>^{*}\notin\gamma^{D}_k$\cr}
\end{equation}
and $\gamma^{D}_k$ is an arbitrarily specified one-cycle in the $k$-th
homology class of the dual lattice.

For the planar lattice the model
breaks down into
two decoupled simple Ising models.
The critical points are two lines $J=J^{*}$ and
$J^{'}=J^{*}$ in the $(J,J^{'})$
plane.

\subsection{The limit $J^{'}\rightarrow 0$}

Here the spin $S^{'}_i$ are decoupled.
For the dual model $J^{'}_D\rightarrow \infty$
(or $J_D\rightarrow \infty$). Therefore
$V_{mn}\tau^{'}_m \tau^{'}_n\equiv 1$, i.e.
$V_{mn}=\tau^{'}_m \tau^{'}_n$. Now $\{V_{mn}\}$
 are pure gauges and $V_i\equiv 1$. Therefore the dual relation
 (\ref{dualz}) becomes
\begin{equation}
\begin{array}{c}{\displaystyle
  \sum_{ \{S_i,U_{ij}\}   }
     \exp \left\{\sum_{<ij>\in L^1 } JU_{ij} S_i S_j
    +\sum_{m\in L^2 } \beta U_m\right\} \hspace{2cm} }\\
{\displaystyle
 \qquad\qquad\qquad = \hfill c \sum_{\{\tau_m\}}
     \exp \left\{\sum_{<mn>\in L_D^1 }
            J_D  \tau_m \tau_n
    +\sum_{m\in L_D^0 } \lambda_D  \tau_m \right\}
}\end{array}
\end{equation}
with $c= 2^{N+B/2-P/2} (\sinh   2J )^{B/2}
      (\sinh   2 \beta)^{P/2}$
This is the known dual relation discussed by Wegner \cite{WEG71} and
Balian et al \cite{BAL75}.


     \begin{figure}[t]\centering


\vspace{-10.cm}

\setlength{\unitlength}{1cm}
\begin{picture}(10,10)
  \put (5,11.0){$\lambda=\infty,\qquad J+J'=J^*$}
  \put (3.6,10.0){$\beta$}
  \put (5.2,9.3){$J^*$}
  \put (2.9,9.0){$\infty$}
  \put (2.3,8.2){$J^*$}
  \put (9.4,3.6){$\infty$}
  \put (10.1,3.4){$J'$}
  \put (3.6,3){$0$}
  \put (0.1,0.7){$\infty$}
  \put (-0.4,-0.5){$J$}
  \put (7.3,-0.3){$\frac{1}{2}\ln\coth J+\frac{1}{2}\ln\coth J'=J^*$}
\end{picture}

 \vspace{1cm}
\caption{
This is a phase diagram for a planar lattice.
Here the real line corresponds to critical points at $\lambda=0$,
the bold real line corresponds to that at $\lambda=\infty$,
and the squares are those for arbitrary $\lambda$. The phase structure
inside the box is unknown.
}
\label{fig4}
     \end{figure}
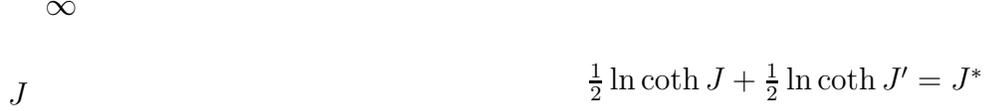

In this case when $\beta\rightarrow \infty$ a phase
transition occurs at $J=J^{*}$.
But note that now $\lambda$ is arbitrary.
Correspondingly when $\lambda_D=0$  and $J^{'}_D \rightarrow \infty$
there are critical points at $J_D=J^{*}$ for arbitrary $\beta_D$.

Including two trivial cases, i.e.
$J=J^{'}=0,\, \beta\rightarrow\infty$ and
$\lambda=0,\, J\rightarrow\infty, \,
\newline J^{'}\rightarrow\infty$ ,
all the critical points discussed above for a planar lattice are plotted in
Fig~\ref{fig4}.
According to the analysis of the mean field method,
inside the box there should be a critical surface.
How to locate this critical surface analytically or
numerically is at present under consideration.

\section {Discussion}

We close the paper by adding some remarks on the Wilson loop.

In the limit $\lambda \rightarrow 0$ and $\beta \rightarrow \infty$,
the Wilson loop along an one-cycle
$\omega$ ( when $\omega$ is a non-boundary one-cycle
the Wilson loop is also called Polyakov string) is
\begin{equation}\begin{array}{c}{\displaystyle
 <\prod_{<ij>\in \omega } U_{ij}>=\hspace{6cm}
}\\{\displaystyle
      \frac{1}{Z_D}\sum^{\alpha-1}_{k=0}
      \sum_{\{\tau_m,\tau^{'}_m\}}
     \exp \sum_{<mn>\in L_D^1 }
            \left(J_D {\tilde V}_{mn} (\gamma_k) \tau_m \tau_n
     +J_D^{'} {\tilde V}_{mn} (\gamma_k+\omega)\tau^{'}_m
    \tau^{'}_n\right)
}\end{array}\end{equation}
The Wilson loop is {\em a topological invariant}, i.e. its value
only depends on the homological class of $\omega$.
This is because when a boundary one-cycle is added to
$\omega$, it can be absorbed into $\{\tau^{'}_m\}$ and the result
is unchanged.
Particularly, when $\omega$ is
a boundary one-cycle, the Wilson loop is equal to one.

Let us further take the limit $J\rightarrow 0$
$(J_D \rightarrow \infty)$. Then

\begin{equation}
 <\prod_{<ij>\in \omega } U_{ij}>=
      \frac{1}{Z_D}
      \sum_{\{\tau^{'}_m\}}
     \exp \sum_{<mn>\in L_D^1 }
            J_D^{'} {\tilde V}^k_{mn} (\omega)\tau^{'}_m \tau^{'}_n
\end{equation}
Consider a $L_1\times L_2$ regular square lattice with periodic
boundary conditions.
The lattice is topologically equivalent to a torus, i.e. $g=1$.


     \begin{figure}[t]\centering


\vspace{-10cm}

\setlength{\unitlength}{1cm}
\begin{picture}(11,10)
\put(3.9,-0.4){$\gamma_1$}
\put(7.0,-0.4){$\tilde\gamma_1$}
\end{picture}

\vspace{1cm}
\caption{
This is a regular torus.
$\gamma_1$ and $\tilde \gamma_1$ devide
the torus into two parts, and inside each part
the spins have the same sign but the two parts are in
opposite signs.}
\label{fig5}
     \end{figure}

In the limit $J^{'}_D \rightarrow 0$, everything is frozen and
$<\prod_{<ij>\in \omega } U_{ij}>$ is trivially equal
to one. However,
suppose $J^{'}_D$ is very large,
for a Polyakov string $\omega=\gamma_1$
the configuration shown in Fig~\ref{fig5}
is dominating, where
$\tilde \gamma_1$ can freely move. Therefore

\begin{equation}
 <\prod_{<ij>\in \gamma_1 } U_{ij}> \sim L_1 e^{-2L_2J^{'}_D}
\end{equation}
This implies the existence of topological excitations.

\vskip 0.7truecm

\newpage

\noindent{\Large\bf Acknowledgement}

\vskip 0.4truecm

One of the authors (Z.B.L) would like to thank the
the Alexander-von-Humboldt-Stiftung for a fellowship.

\end{document}